\documentstyle[12pt,amssymb,epsf]{article}

\makeatletter
\@addtoreset{equation}{section}
\makeatother

\def\ben{\begin{equation}}
\def\een{\end{equation}}

\let\a=\alpha \let\b=\beta   
    
  \let\n=\nu

\let\C=\Chi

\def\nn{\nonumber} \def\bd{\begin{document}} \def\ed{\end{document}}
\def\ds{\documentstyle} \let\fr=\frac \let\bl=\bigl \let\br=\bigr
\let\Br=\Bigr \let\Bl=\Bigl
\let\bm=\bibitem
\let\na=\nabla
\let\pa=\partial \let\ov=\overline
\newcommand{\be}{\begin{equation}}
\newcommand{\ee}{\end{equation}}
\def\ba{\begin{array}}
\def\ea{\end{array}}
\def\ft#1#2{{\textstyle{{\scriptstyle #1}\over {\scriptstyle #2}}}}
\def\fft#1#2{{#1 \over #2}}
\def\del{\partial}
\def\vp{\varphi}
\def\sst#1{{\scriptscriptstyle #1}}
\def\oneone{\rlap 1\mkern4mu{\rm l}}
\def\td{\tilde}
\def\wtd{\widetilde}
\def\ie{\rm i.e.\ }
\def\dalemb#1#2{{\vbox{\hrule height .#2pt
        \hbox{\vrule width.#2pt height#1pt \kern#1pt
                \vrule width.#2pt}
        \hrule height.#2pt}}}
\def\square{\mathord{\dalemb{6.8}{7}\hbox{\hskip1pt}}}
\newcommand{\ho}[1]{$\, ^{#1}$}
\newcommand{\hoch}[1]{$\, ^{#1}$}
\newcommand{\bea}{\begin{eqnarray}}
\newcommand{\eea}{\end{eqnarray}}
\newcommand{\ra}{\rightarrow}
\newcommand{\lra}{\longrightarrow}
\newcommand{\Lra}{\Leftrightarrow}
\newcommand{\ap}{\alpha^\prime}
\newcommand{\bp}{\tilde \beta^\prime}
\newcommand{\tr}{{\rm tr} }
\newcommand{\Tr}{{\rm Tr} }
\def\0{{\sst{(0)}}}
\def\1{{\sst{(1)}}}
\def\2{{\sst{(2)}}}
\def\3{{\sst{(3)}}}
\def\4{{\sst{(4)}}}
\def\5{{\sst{(5)}}}
\def\6{{\sst{(6)}}}
\def\7{{\sst{(7)}}}
\def\8{{\sst{(8)}}}
\def\n{{\sst{(n)}}}
\def\cA{{{\cal A}}}
\def\cB{{{\cal B}}}
\def\cF{{{\cal F}}}
\def\tV{\widetilde V}
\def\tW{\widetilde W}
\def\tH{\widetilde H}
\def\tE{\widetilde E}
\def\tF{\widetilde F}
\def\tA{\widetilde A}
\def\im{{{\rm i}}}
\def\tY{{{\wtd Y}}}
\def\ep{{\epsilon}}
\def\vep{{\varepsilon}}
\def\R{\rlap{\rm I}\mkern3mu{\rm R}}
\def\bD{{{\bar D}}}

\def\R{\rlap{\rm I}\mkern3mu{\rm R}}
\def\bD{{{\bar D}}}
\def\R{{{\Bbb R}}}
\def\C{{{\Bbb C}}}
\def\H{{{\Bbb H}}}
\def\CP{{{\Bbb C}{\Bbb P}}}
\def\RP{{{\Bbb R}{\Bbb P}}}
\def\Z{{{\Bbb Z}}}
\def\bA{{{\Bbb A}}}
\def\bB{{{\Bbb B}}}
\def\bC{{{\Bbb C}}}
\def\bD{{{\Bbb D}}}
\def\bE{{{\Bbb E}}}
\def\bZ{{{\Bbb Z}}}
\def\Re{{{\frak{Re}}}}
\def\Im{{{\frak{Im}}}}
\def\cosec{{\,\hbox{cosec}\,}}
\def\Gm{{\Gamma_{\!\! -}}}
\def\Gp{{\Gamma_{\!\! +}}}
\def\stan{{standard }}
\def\nonstan{{supernumerary }}

\newcommand{\tamphys}{\it Center for Theoretical Physics,
Texas A\&M University, College Station, TX 77843}

\newcommand{\upenn}{\it Department of Physics and Astronomy,\\ University
of Pennsylvania, Philadelphia, PA 19104}

\newcommand{\brussels}{\it Physique Th\'eorique et Math\'ematique,
Universit\'e Libre de Bruxelles,\\ Campus Plaine C.P. 231, B-1050
Bruxelles, Belgium}

\newcommand{\auth}{H. L\"u\hoch{\ast1} and
J.F. V\'azquez-Poritz\hoch{\dagger2}}

\begin{document}
\begin{flushright}

MIFP-03-11\ \ \ \ \
ULB-TH/03-20\\
{\bf hep-th/0305250}\\
May\  2003
\end{flushright}


\begin{center}

{\large {\bf From de Sitter to de Sitter}}

\vspace{20pt}
\auth

\vspace{20pt} {\hoch{\ast}\it George P. and Cynthia W. Mitchell
Institute for Fundamental Physics,\\ Texas A\& M University,
College Station, TX 77843-4242, USA}

\vspace{10pt} {\hoch{\dagger}\brussels}

\vspace{30pt}

\underline{ABSTRACT}
\end{center}

     We obtain $D=6$, ${\cal N}=(1,1$) de Sitter supergravity from a
hyperbolic reduction of the massive type IIA$^*$ theory. We construct
a smooth cosmological solution in which the co-moving time runs from
an infinite past, which is dS$_4\times S^2$, to an infinite future,
which is a dS$_6$-type spacetime with the boundary $R^3\times S^2$.
This provides an effective four-dimensional cosmological model with
two compact extra dimensions forming an $S^2$. Interestingly enough,
although the solution is time-dependent, it arises from a first-order
system {\it via} a superpotential construction. We lift the solutions
back to $D=10$, and in particular obtain two smooth embeddings of
dS$_4$ in massive type IIA$^*$, with the internal space being either
$H^4\times S^2$ or an $H^4$ bundle over $S^2$.  We also obtain the
analogous $D=5$ and $D=4$ solutions.  We show that there exist
cosmological solutions that describe an expanding universe with the
expansion rate significantly larger in the past than in the future.

{\vfill\leftline{}\vfill \vskip 10pt \footnoterule {\footnotesize
\hoch{1} Research supported in part by DOE grant
DE-FG03-95ER40917.

{\footnotesize \hoch{2} Research supported in part by the Francqui
Foundation (Belgium), the Actions de Recherche \phantom{of the}
Concert{\'e}es of the Direction de la Recherche Scientifique -
Communaut\'e Francaise de Belgique, \phantom{of the} IISN-Belgium
(convention 4.4505.86) and by a ``Pole d'Attraction
Interuniversitaire.''}} \vskip -12pt}  \pagebreak
\setcounter{page}{1}

\tableofcontents
\addtocontents{toc}{\protect\setcounter{tocdepth}{2}}
\newpage

\section{Introduction}

      Target spacetime duality \cite{buscher} is the most established
duality in string theory.  It states that a string on a circle with
radius $R$ is equivalent to the same, or another, string theory on a
circle with radius $2\pi\alpha'/R$.  For the heterotic string theory,
the duality was extended to a time-like circle \cite{moore}. However,
as was first observed in \cite{eucl}, the T-duality that relates the
type IIA/IIB theories breaks down in a time-like direction, due to the
Ramond-Ramond fields. Specifically, when reduced on a time-like
circle, the kinetic terms for the R-R fields have opposite signs in
the type IIA and type IIB effective field theories. In order to extend
time-like T-duality to the type II theories, type IIA$^*$ and type
IIB$^*$ theories were introduced in \cite{hull1,hull2}, where the
kinetic terms of the R-R fields have the opposite sign to those in the
usual type II theories.  In particular, the proposed time-like
T-duality relates type IIA to type IIB$^*$ and type IIB to type
IIA$^*$. Lifting type IIA$^*$ to eleven dimensions leads to an
M$^*$-theory which has signature $(2,9)$.

      Just as anti-de Sitter spacetimes arise naturally in type IIB
and M-theories, so de Sitter spacetimes arise in the * theories. The
advantage of such embeddings of dS is that the * theories can still
be viewed as ``supersymmetric,'' at a price that the anti-commutators
of the super-charges are no longer positive definite.  In this paper,
we continue referring this as supersymmetry.  De Sitter supergravities
as hyperbolic reductions of type IIB$^*$ and M$^*$-theories have
recently been obtained in \cite{liu}. In this paper, we consider the
massive type IIA$^*$ theory.  We show that there is a smooth solution
in which the metric is a warped product of dS$_6$ and hyperbolic
4-space $H^4$.  This enables us to perform a Kaluza-Klein hyperbolic
reduction to an ${\cal N}=(1,1)$ gauged supergravity that admits
dS$_6$ as a vacuum solution.  The gauge group is $SU(2)\times U(1)$,
which is the same as that of AdS$_6$ gauged supergravity.  However,
the signs of the kinetic terms of the gauge fields in dS supergravity
are opposite to those in AdS supergravity.  The dilaton and the 3-form
field strength have the standard sign for their kinetic terms.

      Our purpose in studying such a theory is to construct
cosmological solutions.  Recent experimental evidence suggests that
our universe might be de Sitter \cite{exp1,exp2}.  We obtain a smooth
six-dimensional cosmological solution in which the co-moving time runs
from an infinite past, which is dS$_4\times S^2$, to an infinite
future, which is a dS$_6$-type of geometry with the boundary
$R^3\times S^2$.  The Hubble constant in the infinite past is of the
same order of magnitude as that in the infinite future, but the value
decreases.  The increased value of the Hubble constant in the past is
due to the contribution from the 2-form flux. This solution provides
an effective four-dimensional cosmological model, with two extra
compact dimensions forming an $S^2$.

    The cosmological flow is analogous to BPS domain walls in gauged
supergravities, for which there is a renormalization group (RG) flow
\cite{freedman} in the context of the AdS/CFT correspondence
\cite{maldacena,gubser,witten}. In particular, RG flows in massive
type IIA supergravity were constructed in \cite{tuan,sorin2}.  These
are the AdS analogs of the solutions discussed in the present paper.
Cosmological RG flows have recently been discussed in \cite{cvetic}.
However, at least in the truncation or approximation considered, the
cosmology flows from de Sitter to a singularity.  In fact,
cosmological solutions in string and M-theory were constructed
sometime ago \cite{mukherji,low}, which are now known as S-branes
\cite{strom}.  These solutions are typicaly singular.  It has been
argued that one way to resolve the singularity may be for the
cosmological flow to connect early and late time de Sitter
spacetimes. In a general sense, our solution is an example of such a
smooth flow, albeit the division of the early spacetime between dS$_4$
and $S^2$. In light of the proposed dS/CFT correspondence
\cite{park1,park2,strominger1,strominger2,vijay}, the cosmological
evolution that we presently study corresponds to an RG flow from a
three-dimensional Euclidean CFT to a five-dimensional Euclidean CFT
with two compact dimensions on $S^2$.

     Since the reduction ansatz from type IIA$^*$ is consistent, it
allows us to lift these supergravity solutions back to $D=10$.  In
particular, we find two smooth embeddings of dS$_4$ in massive type
IIA$^*$, where the internal space is either $H^4\times S^2$ or an
$H^4$ bundle over $S^2$.

        We also obtain the analogous $D=5$ and $D=4$ de Sitter
supergravity solutions.  An interesting feature of these solutions is
that, although they are time-dependent, they arise from first-order
equations {\it via} a superpotential construction. These solutions are
supersymmetric from the point of view of * theories.  A superpotential
has also been used to construct the analogous solutions in AdS
supergravity which radially interpolate between AdS$_{D-2}\times S^2$
and an AdS$_D$-type geometry with the boundary M$_{D-3}\times S^2$,
where M$_d$ is $d$-dimensional Minkowski spacetime
\cite{sorin1,sorin2}.

       We also demonstrate that there is a larger class of
non-supersymmetric cosmological solutions that arise from the
second-order equations of motion.  For appropriate choices of
parameters, the solution can describe an expanding universe whose
expansion rate is significantly larger in the past than in the future,
providing a realistic inflationary model, with no singularity.
Solutions with such a property can also arise from first-order system
when appropriate matter fields are coupled.

      This paper is organized as follows.  In section 2, we consider
$D$-dimensional de Sitter Einstein-Maxwell gravity as a toy model. We
give a detailed presentation of the construction of a cosmological
solution using a superpotential approach and discuss the properties of
the solutions. In section 3, we consider massive type IIA$^*$ theory
in $D=10$. We perform a Kaluza-Klein hyperbolic reduction to obtain
$D=6$, ${\cal N}=(1,1)$ de Sitter supergravity, obtain a cosmological
solution and analyse its properties. In section 4, we obtain the
analogous solutions in $D=5$ and $D=4$.  We conclude our paper in
section 5.

\section{Cosmological solution in Einstein-Maxwell\\ de Sitter  gravity}

     The Lagrangian for $D$-dimensional Einstein-Maxwell de Sitter gravity
is given by
\be
e^{-1}{\cal L} = R - \ft14\, \eta\, F_\2^2 - (D-1)(D-2)\, g^2\,,
\ee
where $\eta=1$ for the standard kinetic term and $\eta=-1$ for the
kinetic term with opposite sign.  Kinetic terms with ``wrong''
signs occur in type IIA$^*$, type IIB$^*$ and M$^*$-theories, as
we have discussed in the introduction.  We consider a cosmological
solution with the ansatz
\bea
ds^2 &=& -d\tau^2 + a^2\, dx^idx^i + b^2\, d\Omega_2^2\,,\nn\\
F_\2 &=& \lambda\, \Omega_\2\,,\label{cosans}
\eea
where $d\Omega_2^2$ is the metric on a unit 2-sphere $S^2$, 2-torus
$T^2$ or hyperbolic 2-plane $H^2$, and $\Omega_\2$ is the
corresponding volume 2-form.  The functions $a$ and $b$ depend only on
the co-moving time coordinate $\tau$. The equations of motion are
given by
\bea \fft{2\ddot b}{b} + \fft{(D-3)\,\ddot a}{a} &=& (D-1)\, g^2 -
\fft{\eta\, \lambda^2}{2(D-2)\,b^4}\,,\nn\\
\fft{\ddot a}{a} + \fft{2\dot a\dot b}{a\, b} + \fft{(D-4)\,\dot
a^2}{a^2} &=& (D-1)\, g^2
- \fft{\eta\,\lambda^2}{2(D-2)\, b^4}\,,\nn\\
\fft{\ddot b}{b} + \fft{\dot b^2}{b^2} + \fft{(D-3)\, \dot a\,\dot
b}{a\, b} + \fft{\epsilon}{b^2} &=& (D-1)\, g^2 + 
\fft{(D-3)\, \eta\,\lambda^2}{2(D-2)\, b^4}\,, \label{emeom}
\eea
where a dot denotes a derivative with respect to $\tau$ and
$\epsilon=1,0,-1$ for $S^2$, $T^2$ and $H^2$, respectively.  Defining
$a=e^{\varphi_1}$ and $b=e^{\varphi_2}$, we find that the equations
(\ref{emeom}) can be obtained from a Hamiltonian ${\cal H}=T + U=0$,
where
\be T=\ft12\, g_{\alpha\beta}\, (\varphi^\a)'\,
(\varphi^\beta)'\,,\quad\hbox{with}\quad
g_{\a\beta}=\pmatrix{-2(D-3)(D-4) & -4(D-3)\cr
                      -4(D-3) & - 4 \cr}\,,
\ee
and a prime denotes a derivative with respect to $t$, defined by
$d\tau=b^2\, a^{D-3}\, dt$.  The potential $V$ is given by
\be
U=a^{2(D-3)}\, \Big(\ft12\eta\, \lambda^2 - 2\epsilon\, b^2 +
(D-2)(D-1)\, g^2\, b^4\Big)\,. \ee
We find that the potential $U$ can be expressed in terms of a
superpotential $W$ by
\be
U=-\ft12 g^{\alpha\beta}\, \fft{\del W}{\del\varphi^\a}\,
\fft{\del W}{\del\varphi^\beta}\,,
\ee
provided that the following constraint is satisfied
\be
\lambda^2 = -\fft{2\epsilon^2}{\eta\, (D-2)(D-3)\, g^2}\,.
\label{emcon}
\ee
The superpotential $W$ is given by
\be
W=a^{D-3}\, \Big( \fft{2\epsilon}{(D-3)\, g} - 2(D-2)\, g^2\, b^2\Big)
\,.
\ee
The second-order equations (\ref{emeom}) are satisfied if the
first-order equations\\ $(\varphi^\a)'=\pm g^{\a\b}\partial_{\b}W$ are
satisfied. Thus, we arrive at the first-order equations
\be
\fft{\dot a}{a} = g + \fft{\ep}{(D-2)(D-3) g\, b^2}\,,\qquad
\fft{\dot b}{b} = g - \fft{\ep}{(D-2)\, g\, b^2}
\ee
The equations can be solved explicitly, giving a cosmological solution
\bea
ds^2 &=& - \fft{dt^2}{(g\, t)^2\, H^2} + (g\, t)^2\, H^{\ft{D-2}{D-3}}\,
dx^i\, dx^i + t^2\, d\Omega_2^2\,,\nn\\
F_2 &=& \lambda\, \Omega_\2\,,\qquad
H=1 - \fft{\epsilon}{(D-2)\, (g\, t)^2}\,,
\eea
where the charge $\lambda$ is given by
\be
\lambda=\fft{\epsilon}{g\, \sqrt{-\ft12\eta\, (D-2)(D-3)}}\,.
\ee
The solution can also be obtained from the analytical continuation
of the magnetic brane solutions constructed in \cite{sabra1,sabra2}.  It 
is clear that the reality of $\lambda$ requires that $\eta=-1$, implying
that the kinetic term for $F_\2$ must have the ``wrong'' sign
\footnote{In contrast to these solutions, de Sitter black holes
require the standard sign for the $F_\2$ kinetic term \cite{liu}. As
we will see, certain hyperbolic reductions from * theories contain
kinetic terms of both signs and can thereby accommodate both types of
solutions.}. As we have mentioned in the introduction, the * theories
introduced by time-like T-duality precisely have the opposite signs
for the kinetic terms of the R-R fields.  This implies that the
aforementioned type of cosmological solution might arise naturally
from a * theory.

      The solution can also be analytically expressed in terms of the
co-moving coordinate $\tau$ as
\bea
ds^2 &=& -d\tau^2 + e^{\ft{2(D-2)\,g\,\tau}{D-3}}\,
\Big[e^{2g\,\tau} + \fft{\epsilon}{(D-2)\,g^2}\Big]^{-\ft1{D-3}}
\, dx^i\, dx^i\nn\\
&&\qquad\quad +\Big(e^{2g\,\tau} +
\fft{\epsilon}{(D-2)\,g^2}\Big)\, d\Omega_2^2\,.\label{emsol}
\eea
For $\epsilon=0$, the solution is nothing but de Sitter spacetime in
$D$-dimensions.  For $\epsilon=-1$, the solution approaches a
dS$_D$-type spacetime in the future but has a naked singularity at a
certain time in the past.  We are interested in the solution with
$\epsilon=1$.  In this case, the solution is regular everywhere and
time runs from an infinite past, which is dS$_{D-2}\times S^2$ with
Hubble constant $H=\ft{D-2}{D-3}\,g$, to an infinite future, which
is a dS$_D$-type spacetime with Hubble constant $H=g$.  Thus we
see that the Hubble constant at infinite past is of the same magnitude
as that at the infinite future.  The value decreases by
$\ft{100}{D-2}\, \%$.  The larger value of $H$ in the past is due to
the contribution from the 2-form flux.

       Note that for $\epsilon=1$, the dS$_4\times S^2$ of the
infinite past is itself a solution. It can be obtained by taking
$\tau\rightarrow \tau + \tau_0$ and sending $\tau_0$ to $-\infty$.
With an appropriate rescaling of $x^i$, the metric is given by
\be
ds^2=-d\tau^2 + e^{\ft{2(D-2)\,g\,\tau}{D-3}}\, dx^i\, dx^i + 
\fft{1}{(D-2)\, g^2}\, d\Omega_2^2
\ee

      It is worth mentioning that the second-order equations
(\ref{emeom}) admit a larger class of cosmological solutions that run
from dS$_{D-2}$ in the past to dS$_D$ in the future, regardless of the
sign of $\epsilon$ and $\eta$.  To see this, we first consider the
``fixed-point'' solution dS$_{D-2}\times S^2$ or dS$_{D-2}\times
H^2$, where $b$ is a constant.  The solution is given by
\bea
ds^2 &=& -d\tau^2 + e^{2\gamma\, \tau}\, dx^i\, dx^i + 
b^2\, d\Omega_2^2\,,\nn\\
b^2 &=& \fft{1}{2(D-1)\, g^2}\,\Big(\epsilon \pm \sqrt{\epsilon^2 -
\fft{2(D-1)(D-3)\, g^2\, \eta\, \lambda^2}{D-2}}\,\Big)\,,\nn\\
\gamma^2 &=& \fft{(D-1)(D-2)\, g^2}{(D-3)^2} - 
\fft{\epsilon}{(D-3)^2\, b^2}\,.\label{2ndsol}
\eea
Thus for $\eta=1$, the value of $\lambda$ is restricted to satisfy
$g^2\, \lambda^2 \le (D-2)/(2(D-1)(D-3))$.  For $\eta=-1$, there is no
such restriction; in this case, with $\epsilon=1$, when $\lambda$
satisfies the constraint (\ref{emcon}), the solution becomes the one
that can arise from the first-order system, as we discussed earlier.
It is straightforward to demonstrate using numerical methods that
there exist cosmological solutions that run from (\ref{2ndsol}) in the
infinite past to a dS$_D$-like spacetime in the infinite future.

      In order to construct a realistic model, we would like to have
the effective cosmological constant much larger in the past than
in the future.  This can be achieved by the choice of $\epsilon=-1$
and $\eta=-1$, with $g^2\, \lambda^2 <<1$.  The resulting
dS$_{D-2}\times H^2$ metric in the infinite past is given by
(\ref{2ndsol}), but with $b$ and $\gamma$ given by
\be
\gamma^2=\fft{(D-1)(D-2)\,g^2}{(D-3)^2} + \fft{2(D-2)}{(D-3)^3\,
\lambda^2}\,,\qquad
b^2= \fft{(D-3)\lambda^2}{2(D-2)}\,.
\ee
Thus the Hubble constant $H\sim 1/\lambda$ in the past for small
$\lambda$ can be significantly larger than $H=g$, the Hubble constant
in the future. This is analogous to the recent observation that a
vacuum solution dimensionally-reduced on a compact hyperbolic manifold
of time-varying volume might yield an inflationary cosmology
\cite{townsend}, which is a special case of an S-brane solution 
\cite{mukherji,ohta1}.

     Clearly, the charge $\lambda$ is restricted to take the value
defined in (\ref{emcon}) in solution that arises from the first-order
system.  Thus the Hubble constant in the future and in the past are of
the same orders of magnitude.  In the general solutions from the
second-order equations, such a restriction no longer applies, and
therefore we can adjust parameters in such a way that the expansion
rate of the universe in the past is significantly larger than that in
the future, in line with observational data.  In fact, as we shall
show in the next section, when appropriate matter fields are coupled,
even the solutions arising from the first-order system associated with
supersymmetry can have expansion rate significantly larger in the past
than in the future.

      Although here we considered the general de Sitter
Einstein-Maxwell supergravity, only the $D=4$ case can be embedded in
the * theories.  In other dimensions, the de Sitter supergravity
coming from the reduction of * theories involves additional fields
such as a dilaton and antisymmetric tensor fields. We shall discuss
this with further detail in subsequent sections.  Nevertheless, the
solution (\ref{emsol}) captures the essence of all the cosmological
solutions we obtain in this section, namely the metric is
dS$_{D-2}\times S^2$ in the infinite past and of dS$_D$-type in the
infinite future.  The contributions of scalars or additional field
strengths might alter the Hubble constant in the past but only in a
mild way.

\section{$D=6$ cosmological solution}

\subsection{$D=6$ de Sitter supergravity from massive type IIA$^*$}

        Massive type IIA supergravity was constructed in
\cite{romans}.  The bosonic Lagrangian for the massive type IIA$^*$
supergravity can be obtained by changing the sign of the kinetic terms
for the R-R fields, namely the 2-form and 4-form field strengths,
together with a sign change for the cosmological term; it is given by
\bea
{\cal L}_{10}\!\!\! &=&\!\!\!\hat R\, {\hat *\oneone} -
\ft12 {\hat *d\hat \phi}\wedge d\hat \phi
+ \ft12 e^{\fft32\hat\phi}\, {\hat *\hat F_\2}\wedge \hat F_\2 -
\ft12 e^{-\hat \phi}\, {\hat *\hat F_\3}\wedge \hat F_\3+
\ft12 e^{\fft12\hat\phi}\, {\hat *\hat F_\4}\wedge \hat F_\4 \nn\\
\!\!\!&&\!\!\!
+\ft12 d\hat A_\3\wedge d\hat A_\3 \wedge \hat A_\2 + \ft16 m\,
d\hat A_\3 \wedge (\hat A_\2)^3
 +\ft1{40} m^2 (\hat A_\2)^5 +\ft12 m^2 e^{\fft52\hat \phi}\,
{\hat *\oneone},\label{romans1} \eea
where the field strengths are given in terms of potentials by
\bea
\hat F_\2 &=& d\hat A_\1 + m\, \hat A_\2\ ,\qquad \hat F_\3 =
d\hat A_\2\,,\nn\\
\hat F_\4 &=& d\hat A_\3 + \hat A_\1\wedge d\hat A_\2 + \ft12 m\,
\hat A_\2\wedge \hat A_\2\,.\label{romfields}
\eea
Here we present the Lagrangian in the language of differential forms,
as in \cite{massiveab}.  The theory admits a solution in the form of a
warped product of dS$_6\times H^4$, given by
\bea
ds_{10}^2 &=& (\cosh\xi)^{\ft1{12}}\, \Big[
ds_{\sst{dS_6}}^2 + 2g^{-2}\, \Big(d\xi^2 + 
\ft14\sinh^2\xi\, (\sigma_1^2
+ \sigma_2^2 + \sigma_3^2)\Big)\Big]\,,\label{ds6s4}\\
\hat F_\4 &=& \ft{5\sqrt2}{6}\, g^{-3} (\cosh\xi)^{1/3}\,
\sinh^3\xi\, d\xi\, 
\wedge \sigma_1\wedge\sigma_2\wedge\sigma_3\,,\qquad
e^{\hat \phi}=(\cosh\xi)^{-5/6}\,,\nn
\eea
where $\sigma_i$ are $SU(2)$ left-invariant 1-forms which satisfy
$d\sigma^i = -\ft12 \ep_{ijk}\, \sigma^j\wedge \sigma^k$. In other
words, the metric for the hyperbolic $H^4$ is written as a foliation
by $S^3$.  This solution is an analytical continuation of the warped
product of AdS$_6$ and $S^4$ obtained in \cite{brandoz}. It is of
interest to note that, while the warped product of AdS$_6$ and $S^4$
is singular at the equator of the $S^4$, our solution (\ref{ds6s4}) is
regular everywhere.

     We can now consider the Kaluza-Klein hyperbolic reduction of the
massive type IIA$^*$ theory. The $S^4$ reduction of the usual massive
type IIA supergravity was obtained in \cite{massive}. Here we need
only to perform an analytic continuation of the reduction ans\"atze
obtained in \cite{massive}, which is then given by
\bea
d\hat s_{10}^2&=&
(\cosh\xi)^{\fft1{12}}\, X^{\fft18}\Big[
\Delta^{\fft38}\, ds_6^2 + 2g^{-2}\, \Delta^{\fft38}\, X^2\, d\xi^2
+\ft12g^{-2}\, \Delta^{-\fft58}\, X^{-1}\, \sinh^2\xi
\sum_{i=1}^3 (h^i)^2\Big]\,,\nn\\
\hat F_\4 &=& -\ft{\sqrt2}{6}\, g^{-3}\, c^{1/3}\, s^3\, \Delta^{-2}\,
U\, d\xi\wedge\ep_\3 -\sqrt2 g^{-3}\, c^{4/3}\, s^4\, \Delta^{-2}\,
X^{-3}\, dX\wedge \ep_\3 \nn\\
&&-\sqrt2 g^{-1}\,
c^{1/3}\, s\, X^4\, {*F_\3}\wedge d\xi
-\ft1{\sqrt2} c^{4/3}\, X^{-2}\, {*F_\2} \nn\\
&& +\ft1{\sqrt2} g^{-2}\,
c^{1/3}\, s\, F_\2^i \, h^i\wedge d\xi -\ft1{4\sqrt2} g^{-2}\,
c^{4/3}\, s^2\, \Delta^{-1}\, X^{-3}\,  F_\2^i \wedge
h^j\wedge h^k\, \ep_{ijk}\,,\label{fans}\\
\hat F_\3 &=& c^{2/3}\, F_\3 + g^{-1}\, c^{-1/3}\, s\, F_\2\wedge d\xi
\,,\nn\\
\hat F_\2 &=& \ft1{\sqrt2}\, c^{2/3}\, F_\2\,,\qquad
e^{\hat\phi} = c^{-5/6}\, \Delta^{1/4}\, X^{-5/4}\,,\nn
\eea
where $X$ is related to the dilaton $\phi$ by
$X=e^{-\fft1{2\sqrt2}\phi}$ and
\bea
\Delta &\equiv & -X\sinh^2\xi +X^{-3} \cosh^2 \xi\,,\nn\\
U &\equiv& X^{-6}\, c^2 + 3 X^2\, s^2 - 4 X^{-2}\, s^2 - 6 X^{-2}\,.
\label{deltau}
\eea
We have defined $h^i\equiv \sigma^i+g\, A_\1^i$, $\ep_\3\equiv
h^1\wedge h^2\wedge h^3$, $s=\sin\xi$ and $c=\cos\xi$.  Also, $*$ is
the six-dimensional Hodge dual.  (There should be no confusion between
Hodge dual $*$ and * theories.) The gauge coupling constant $g$ is
related to the mass parameter $m$ of the massive type IIA theory by
$m= \ft{\sqrt2}{3}\, g$.  The resulting six-dimensional theory is
given by
\bea
{\cal L}_6 &=& R\, {*\oneone} -\ft12 {*d\phi}\wedge d\phi
+ g^2\Big(\ft29 e^{\fft{3}{\sqrt2}\phi} -\ft83 e^{\fft1{\sqrt2}\phi} -
2 e^{-\fft1{\sqrt2}\phi}\Big)\,  {*\oneone}\nn\\
&&-\ft12 e^{-\sqrt2\phi}\, {*F_\3\wedge F_\3} +\ft12
e^{\fft1{\sqrt2}\phi}\, \Big( {*F_\2}\wedge F_\2 + {*F_\2^i}\wedge
F_\2^i \Big) \label{d6lag}\\
&& + A_\2\wedge(\ft12
dA_\1\wedge dA_\1 +\ft13 g\, A_\2\wedge dA_\1 +\ft2{27}
g^2\, A_\2\wedge A_\2 +\ft12 F_\2^i\wedge F_\2^i)\,,\nn
\eea
where $F_\3=dA_\2$, $F_\2= dA_\1 + \ft23g\, A_\2$ and $F_\2^i =
dA_\1^i - \ft12 g\, \ep_{ijk} A_\1^j\wedge A_\1^k$.

       The de Sitter gauged supergravity in $D=6$ which we have
obtained from the hyperbolic reduction of the massive type IIA$^*$
theory was also constructed in $D=6$ directly in \cite{dv}.  It has an
$SU(2)\times U(1)$ gauge symmetry.  This is the same as in the AdS
gauged supergravity in $D=6$.  The difference between the theories is
that the kinetic terms for the gauge fields have opposite signs.
Furthermore, the cosmological constant is positive instead of
negative.  However, as we have shown in section 2, these features are
necessary in order to construct the ``supersymmetric'' cosmological
solution which is presented in the next subsection.

\subsection{Cosmological solution}

     We use the metric, the dilaton and a 2-form
field strength to construct a cosmological solution.  The relevant
Lagrangian is given by
\be
e^{-1} {\cal L}_6 = R - \ft12 (\del\phi)^2 +
\ft14\, e^{\ft1{\sqrt2}}\, F_\2^2 +
g^2\Big(\ft29 e^{\fft{3}{\sqrt2}\phi} -
\ft83 e^{\fft1{\sqrt2}\phi} - 2 e^{-\fft1{\sqrt2}\phi}\Big)\,.
\ee
We consider the same ansatz for the cosmological solution in
(\ref{cosans}).  The system admits the following first-order equations
\bea
\dot \phi &=& \sqrt2\, \Big(\fft{\sqrt2}{4}\epsilon\,g^{-1}\,
e^{\ft{1}{2\sqrt2}\,\phi}\, b^{-2} + \fft{dW}{d\phi}\Big)\,,\nn\\
\fft{\dot b}{b} &=& -\fft1{4\sqrt2}\,\Big( 3\epsilon\,g^{-1}\,
e^{\ft{1}{2\sqrt2}\,\phi}\, b^{-2} + W\Big)\,,\nn\\
\fft{\dot a}{a} &=& \fft1{4\sqrt2}\, (\epsilon\, g^{-1}\,
e^{\ft{1}{2\sqrt2}\,\phi}\, b^{-2} - W\Big)\,,
\eea
provided that the charge $\lambda$ is fixed to be
$\lambda=\epsilon/g$.  The superpotential $W$ is given by
\be
W=-2g\, \Big(e^{-\ft1{2\sqrt2}\,\phi} + \ft13 e^{\ft{3}{2\sqrt2}\,\phi}
\Big)\,.
\ee

     For all $\epsilon$, the solution runs to a dS$_6$-type solution
in the infinite future, with the Hubble constant being
$H=\ft{\sqrt{2}}{3}\,g$.  For $\epsilon=-1$ and 0, the solution has a
naked singularity at a particular time in the finite past.  We are
interested in $\epsilon=1$, for which there is a fixed-point solution,
namely
\bea
ds_6^2 &=&-d\tau^2 + e^{\ft{2}{6^{1/4}}\, g\, \tau}\,
(dx_1^2 + dx_2^2 + dx_3^2) + \sqrt{\ft32}\, g^{-2}
\, d\Omega_2^2\,,\nn\\
e^{\sqrt2\,\phi} &=& \ft32\,,\qquad F_\2=g^{-1}\, \Omega_\2\,.
\label{ds4s2}
\eea
It describes a direct product of dS$_4\times S^2$.  Using a numerical
approach, it is straightforward to demonstrate that there exists a
solution whose co-moving time runs from infinite past, with the above
dS$_4\times S^2$ with the Hubble constant $6^{-1/4}\,g$, to an
infinite future, with a dS$_6$-type metric having boundary $R^3\times
S^2$, and a Hubble constant $\ft{\sqrt{2}}{3}\, g$.  Again, the two
Hubble constants are of the same orders of magnitude but the one in
the infinite future is slightly smaller than the one in the infinite
past.

        As we have discussed in section 2, the reason for the Hubble
constant to take the values of the same magnitude in the past and in
the future is that the first-order equations arising from
supersymmetry require that the charge carried by $F_\2$ take a
specific value.  There also exists a more general class of solutions
of the second-order equations of motion, where this restriction does
not apply.  It is rather straightforward to show, as in section 2,
that by adjusting the parameters appropriately, we can obtain a
realistic inflationary model whose expansion rate in the past is
significantly larger than that in the future.

\subsection{Smooth embedding of dS$_4$ in massive type IIA$^*$}

     Owing to the consistency of the reduction ansatz, the solutions
can all be lifted back to $D=10$. As we have seen earlier, the dS$_6$
becomes a smooth warped product of dS$_6$ and $H^4$.  Here we consider
the oxidation of the dS$_4\times S^2$ solution, given in
(\ref{ds4s2}).  There are two possible endpoints in $D=10$.  If the
2-form field strength is taken to be in the $U(1)$ factor in the
$SU(2)\times U(1)$ gauge group, then there is no non-trivial bundle
structure in $D=10$, and the metric is given by
\bea ds_{10}^2&=&(\ft23)^{\ft{1}{32}}\,({\rm cosh}\,
\xi)^{\ft{1}{12}}\,\Delta^{\ft38}\, \Big[ ds_{{\rm
dS}_4}^2+ \sqrt{\ft32}\, g^{-2}\, d\Omega_2^2\nn\\
&& \qquad +g^{-2}\Big( (\ft83)^{\ft12}\,d\xi^2
+\ft12\,(\ft32)^{\ft14}\, \Delta^{-1}\, {\rm sinh}^2\, \xi\,
d\Omega_3^2\Big) \Big]\,,\label{ds41}
\eea
where
\be \Delta=(\ft32)^{3/4}\, \cosh^2\xi - (\ft23)^{1/4}\,
\sinh^2\xi\ >0\,. \label{deltav}
\ee
The dS$_4$ metric is given by
\be ds_{{\rm
dS}_4}^2=-d\tau^2+e^{\ft{2}{6^{1/4}}\,g\,\tau}\,(dx_1^2
+dx_2^2+dx_3^2)\,. \ee
The resulting ten-dimensional metric is also completely smooth.

    An alternative possibility is that the 2-form field strength in
(\ref{ds4s2}) corresponds to a $U(1)$ subgroup of the $SU(2)$ gauge
group.  If we lift this solution back to $D=10$, then there is a
non-trivial bundle structure in the metric, which is given by
\bea ds_{10}^2&=&(\ft23)^{\ft{1}{32}}\,({\rm cosh}\,
\xi)^{\ft{1}{12}}\,\Delta^{\ft38}\, \Big[ ds_{{\rm
dS}_4}^2+\sqrt{\ft32}\, g^{-2}\, d\Omega_2^2\nn\\
&&\qquad +g^{-2}\Big( (\ft83)^{\ft12}\,d\xi^2
+\ft12\,(\ft32)^{\ft14}\, \Delta^{-1}\, {\rm sinh}^2\, \xi\,
(d\wtd\Omega_2^2 + (d\psi + \bar A_1)^2 \Big) \Big]\,,\label{ds42}
\eea
where $\Delta$ is given by (\ref{deltav}) and
\be
d\bar A_\1=\Omega_\2 + \wtd \Omega_\2\,.
\ee
The internal six-dimensional space can be viewed as an $H^4$
bundle over $S^2$.

     Thus, we have obtained two smooth embeddings of dS$_4$ in massive
type IIA$^*$, one of which has $H^4\times S^2$ as its internal space,
whilst the other is an $H^4$ bundle over $S^2$.  If we lift the
cosmological solution back to $D=10$, then we obtain ten-dimensional
smooth cosmological solutions, which in the infinite past are either
(\ref{ds41}) or (\ref{ds42}) and they approach (\ref{ds6s4}) in the
infinite future.

\subsection{Matter-coupled de Sitter supergravity}

         So far we have considered only the pure ${\cal N}=(1,1)$ de
Sitter supergravity.  It is also possible to add matter
multiplets. Here we are interested in a vector-tensor multiplet, since
the theory then allows us to truncate to $U(1)^2$, with the Lagrangian
given by\footnote{Though in general not the case, the truncation is
consistent for the present purpose of constructing cosmological
solutions.}
\be
\hat e^{-1}{\cal L}_6 = \hat R - \ft12 (\del \phi_1)^2 -
\ft12 (\del \phi_2)^2 -
\hat V +\ft14 \sum_{i=1}^2 X_i^{-2}\, (\hat F_\2^i)^2\,,
\ee
where $X_i=e^{\ft12\vec a_i\cdot \vec \phi}$ with
\be
\vec a_1=(\sqrt2\,, \fft1{\sqrt2})\,,\qquad
\vec a_2=(-\sqrt2\,, \fft1{\sqrt2})\,.
\ee
The scalar potential is given by
\be
\hat V=-\ft49 g^2\, (X_0^2 - 9 X_1\, X_2 - 6 X_0\, X_1 - 6 X_0\, X_2)\,,
\ee
where $X_0=(X_1\,X_2)^{-3/2}$.  The cosmological solution can be
obtained from the following first-order equations
\bea
\dot{\vec \phi} &=& \sqrt2\,\Big(-\fft{\epsilon}{2\sqrt2\,g}\,
(q_1\, \vec a_1\, X_1^{-1} + q_2\, \vec a_2\, X_2^{-1})\,
b^{-2} + \fft{dW}{d\vec \phi}\Big)\,,\nn\\
\fft{\dot b}{b} &=& -\fft{1}{4\sqrt2}\, \Big(\fft3{\sqrt2\,g}\,
\epsilon\,(q_1\, X_1^{-1} + q_2\, X_2^{-1})\, b^{-2} +
W\Big)\,,\nn\\
\fft{\dot a}{a} &=& \fft{1}{4\sqrt2}\, \Big(
\fft{\epsilon}{\sqrt2\,g}\,(q_1\, X_1^{-1} + q_2\, X_2^{-1})\,
b^{-2} - W\Big)\,, \label{6} \eea
provided that the two $U(1)$ charges are given by
$\lambda_i=\epsilon\, g^{-1}\, q_i$ with $q_1+q_2=1$.  Note that
the $(\vec \phi ,b)$ fields form a closed system.  The
superpotential $W$ is given by
\be
W=\fft{g}{\sqrt2}\,(\ft43 X_0 + 2 X_1 + 2 X_2)\,.
\ee
The case with $q_1=q_2=\ft12$ reduces to the previous example of pure
de Sitter gauged supergravity.  For general $q_i$, the solution can be
obtained by analytical continuation of the magnetic brane solutions in
\cite{sorin1,sorin2}.  In particular, if $q_1=1$ and $q_2=0$, then we
can consistently set $\phi_1=2\phi_2$ and (\ref{6}) can be solved
explicitly by making the coordinate transformation
$d\tau=e^{\ft{3}{\sqrt{8}} \phi_2}\, dy$. Defining
\be F\equiv e^{-\sqrt{8}\,\phi_2}\,,\quad G\equiv
e^{\ft{1}{\sqrt{2}}\,\phi_2}\,b^2\,, \ee
we find that the first two equations in (\ref{6}) yield
\be G'-\fft{\epsilon}{g}+\ft43\,g\, G=0\,. \ee
The solution is
\be G=g^{-2}\, (e^{-\ft43\,g\, y}+\ft34\,\epsilon)\,. \ee
We have absorbed a trivial integration constant by a constant
shift in the coordinate $y$.  The dilaton equation of motion gives
\be \fft{F'}{F}=-\fft{\epsilon}{g\,G}+2g\,(1-F^{-1})\,. \ee
Plugging in $G$, we find that
\bea F &=& \frac{e^{-\ft43\, g\, y}+\ft94\, \epsilon+c_1\,
e^{\ft23\, g\, y}}{g^2\, G}=e^{-\sqrt{8}\phi_2}\,,\nn\\
a^2 &=& c_2\, e^{-\ft43\, g\, y}\,
e^{-\ft{1}{\sqrt{2}}\,\phi_2}\,,\quad b^2 =g^{-2}\, (e^{-\ft43\,
g\, y}+\ft34\, \epsilon)\, e^{-\ft{1}{\sqrt{2}}\, \phi_2}\,. \eea
The metric of the solution can be expressed, using the new
coordinate $t=g^{-1}\, e^{-\ft23\, g\, y}$, as
\be ds_6^2 = -\ft94\, H^{\ft34}\, \fft{dt^2}{(g\, t)^2}+ (g\,
t)^2\, H^{-\ft14}\Big(dx_i^2 + (1 +\ft34\, \fft{\epsilon}{(g\,
t)^2})\, g^{-2}\, d\Omega_2^2\Big) \,, \label{singlec} \ee
where the function $H$ and the dilaton are given by
\be e^{\sqrt{8}\, \phi_2}=H=\fft{1 + \ft34\, \fft{\epsilon}{(g\,
t)^2}}{1 + \ft94 \, \fft{\epsilon}{(g\, t)^2} + \fft{c_1}{(g\,
t)^3}}\,.\label{genh} \ee
As shown previously for the case of dS$_4\times S^2$, the above
cosmological solution can be lifted to type IIA* theory.

To understand the solution better, it is instructive to write the
function $H=\wtd H/W$, where the definitions of $\wtd H$ and $W$
can be straightforwardly read off from (\ref{genh}). Then the
metric can be expressed as
\be ds_6^2 = \wtd H^{-\ft14}\, W^{\ft14}\,\Big[-\ft94\, \wtd H\,
W^{-1}\, \fft{dt^2}{(g\, t)^2} +(g\, t)^2\, (dx_i^2 + \wtd H\,
g^{-2}\, d\Omega_2^2) \Big]\,. \label{gendmetricsol} \ee
Thus, the solution can be viewed as an intersection of a spatial
domain wall, characterized by the function $H$, and a brane with a
three-dimensional Euclidean worldvolume\footnote{This is known as an
S2-brane \cite{mukherji,gutperle}.}, characterized by the function
$\wtd H$.

     In the infinite future, $t\rightarrow \infty$, $H\rightarrow 1$
and the metric behaves as
\be ds_6^2=-\ft94\, \fft{dt^2}{(g\, t)^2} +(g\,t)^2\, (dx_i^2+
g^{-2}\, d\Omega_2^2)\,.\label{larged1} \ee
If $\epsilon=0$, in which case $d\Omega_2^2$ is the metric on a
2-torus, then the above metric describes locally dS$_6$ spacetime.
The full solution with $\epsilon=0$ describes a spatial domain wall.
This can be viewed as a distribution of S-branes from the ten or
eleven-dimensional point of view.

      For $\epsilon=\pm 1$, the above metric can be viewed as a
spatial domain wall wrapped on $\Omega^2$. For $\epsilon=1$, the
metric is regular everywhere provided that the constant $c_1=0$.  The
corresponding metric interpolates between dS$_4\times S^2$ in the
infinite past to the dS$_6$-type metric (\ref{larged1}) in the
infinite future. When $c_1$ is non-vanishing, the $c_1$ term can
dominate at early time and hence the metric can become singular.  

    If we set $c_1=0$ and rescale the coordinates as
\be\label{resc} g\,t \rightarrow k\, g\, t\,,\qquad x_i
\rightarrow \ft{1}{k}\, x_i\,, \ee
then for $k\ra 0$ we obtain the geometry dS$_4\times S^2$ with a
constant dilaton. If we instead take $c_1 = k\, \tilde c_1\,
m/\gamma$, then after sending $k\ra 0$ we obtain the solution
\bea ds_6^2 &=& -\ft94\, H^{3/4}\, \fft{dt^2}{(g\, t )^2} +
\fft{(g\,t)^2}{ H^{1/4}}\, dx_i^2 +
\ft34\, \fft{1}{g^2H^{1/4}}\, d\Omega_2^2 \,,\nn\\
H^{-1} &=& 3 - \fft{\td c_1}{g\,t}\,, \label{d6h2} \eea
which approaches dS$_4\times S^2$ in the asymptotic region, where
dS$_4$ approaches the boundary. The early time behavior is dependent
on ${\tilde c}_1$, but always singular. Thus, the dS$_4\times S^2$
solution can exist in either the infinite past or future.  For
$\epsilon=-1$, the solution is also always singular at early times.

\subsection{Large past-future Hubble constant ratio}

      For general values of $q_1+q_2=1$, the first-order equations
(\ref{6}) cannot be solved analytically.  However, it is
straightforward to find the fixed-point solution of dS$_4\times S^2$
or dS$_4\times H^2$ where $b$ and $\vec\phi$ are constants.  The
solution is given by
\bea
&&e^{\sqrt2\,\phi_1}=\ft32\Big(1-q \pm \sqrt{(1-q)^2 +\ft49 q}\Big)\,,
\qquad e^{-\sqrt2\, \phi_2}=\ft32\cosh(\phi_1/\sqrt2)\,,\nn\\
&&b^2=\fft{q_1 + q_2\, e^{\sqrt2\,\phi_1}}{4\epsilon\, g^2}\,,\qquad
a^2={\rm Exp}\Big(\ft13g\, e^{-\ft3{\sqrt8}\, \phi_2}\, \tau\Big)\,,
\label{d6gensol}
\eea
where $q=\lambda_1/\lambda_2$.  Thus, it is clear that if $q\in
[0,\infty]$, we have $\epsilon=1$ corresponding to $S^2$, and for
$q\in (-\infty,0)$, we have $\epsilon=-1$ corresponding to $H^2$.
Using a numerical method, we can verify that there are cosmological
solutions which are (\ref{d6gensol}) in the infinite past and are
of ds$_6$-type in the future.  The ratio of the Hubble constant in the 
infinite past and future can be straightforwardly calculated.  In 
particular, for $-q >> 1$, the ratio is given by
\be
\fft{H_{\rm past}}{H_{\rm future}}\sim
\fft{\sqrt2}{16}\,(\ft32\, \sqrt{-3q})^{3/4}\,,
\ee
which can be arbitrarily large.  It is surprising that we can get a 
somewhat realistic cosmological model from a first-order system, which
implies supersymmetry from the point of view of * theories.

\section{Further cosmological solutions}

\subsection{$D=5$}

   De Sitter supergravity from a hyperbolic 5-space reduction of type
IIB$^*$ theory was obtained in \cite{liu}.  Let us now consider the
minimal gauged supergravity in $D=5$ coupled to two vector
multiplets. The Lagrangian is given by
\be
e^{-1}{\cal L}_5=R - \ft12 (\del \phi_1)^2 -\ft12 (\del
\phi_2)^2 - \fft14\sum_{i=1}^3 X_i^{-2}\,\eta_i\,(F_\2^i)^2 - V +
\ft14 e^{-1}\, \epsilon^{\mu\nu\rho\sigma\lambda}\,
F^1_{\mu\nu}\,F^2_{\rho\sigma}\, A^3_\lambda\,, \ee
with the scalar potential
\be
V=4g^2 \sum_{i=1}^3
         X_i^{-1}\,.\label{stdscalarpot}
\ee
The quantities $X_i$
are given by
\bea
&&X_i=e^{\ft12\vec a_i\cdot\vec \phi}\,,\nn\\
&&\vec a_1=(\sqrt2\,, \fft{2}{\sqrt6})\,, \qquad
\vec a_2=(-\sqrt2\,,\fft2{\sqrt6})\,,\qquad \vec
a_3=(0,- \,\fft4{\sqrt6})\,.
\eea
Note that $\eta_i=\{-,-,+\}$ \cite{liu}.  Thus, the reality condition
does not permit one to construct pure de Sitter Einstein-Maxwell
supergravity. For the present purpose of constructing a cosmological
solution, we select the field strengths $F_\2^1$ and $F_\2^2$, whose
kinetic terms have the ``wrong'' sign. With the metric ansatz of the
form (\ref{cosans}), we find that the system admits the following
first-order equations
\bea
\dot {\vec\phi} &=& \sqrt2\Big(\fft{\epsilon}{2\sqrt2\,g}\,
(q_1\,\vec a_1\, X_1^{-1} + q_2\, \vec a_2\, X_2^{-1})\, b^{-2} +
\fft{dW}{d\vec\phi}\Big)\,,\nn\\
\fft{\dot b}{b} &=& -\fft{1}{3\sqrt2}\, \Big(
-\sqrt2\,\epsilon\, (q_1\, X_1^{-1} + q_2\, X_2^{-1})\, b^{-2} +
W)\,,\nn\\
\fft{\dot a}{a} &=& \fft{1}{3\sqrt2}\, \Big(-\fft{\epsilon}{\sqrt2}\,
(q_1\, X_1^{-1} + m_2\, X_2^{-1})\, b^{-2} - W\Big)\,,
\eea
provided that the charges of $F_\2^1$ and $F_\2^2$ are given by
$\lambda_i= \epsilon\, g^{-1}\, q_i$, with $q_1 +q_2=1$.  Here,
the superpotential $W$ is given by
\be
W=\sqrt2\, g\, \sum_i X_i\,.
\ee

        In order to have a fixed-point solution of dS$_3\times S^2$,
it is necessary to have $q_1=q_2=\ft12$, in which case the solution
is given by
\bea
ds^2 &=& -d\tau^2 + e^{2\,\,2^{2/3}\, g\, \tau}\, (dx_1^2 + dx_2^2) +
\fft{1}{2\,\, 2^{1/3}\, g^2}\, d\Omega_2^2\,,\nn\\
X_1&=&X_2=X_3^{-1/2}= (\ft12)^{1/3}\,.
\eea
It is straightforward, using numerical methods, to show that there
exists a smooth solution that runs from dS$_3\times S^2$ in the
infinite past to an infinite future which has a dS$_5$-type metric
with the boundary being $R^2\times S^2$.

\subsection{$D=4$}

         M$^*$-theory is a (2,9) theory that admits a
dS$_4\times$AdS$_7$ vacuum solution.  Kaluza-Klein reduction on
AdS$_7$ gives rise to a de Sitter gauged supergravity in $D=4$
\cite{liu}.  For the truncation to the $U(1)^4$ subsector, the
bosonic Lagrangian is given by
\be
e^{-1}{\cal L}_4=R - \ft12 (\del \phi_1)^2 -\ft12 (\del
\phi_2)^2 -\ft12 (\del \phi_3)^2 + \fft14\sum_{i=1}^4 X_i^{-2}
(F_\2^i)^2 - \hat V \,, \ee
with the scalar potential
\be \hat V=4g^2 \sum_{i<j} X_i\, X_j\,. \label{d4V} \ee
The quantities $X_i$ are given by
\bea
&&X_i=e^{\ft12\vec a_i\cdot\vec \phi}\,,\nn\\
&&\vec a_1=(1, 1, 1),\ \ \vec a_2=(1, -1, -1),\ \ \vec a_3=(-1, 1,
-1),\ \ \vec a_4=(-1, -1, 1). \eea
Note that all four of the above kinetic terms for the field
strengths $F_\2^i$ have the ``wrong'' sign \cite{liu}.  Thus, this
theory can be truncated to pure de Sitter Einstein-Maxwell
supergravity. The cosmological solution ansatz admits the
first-order equations
\bea
\dot {\vec \phi}&=&
\sqrt2\,\Big(\fft{\epsilon}{2\sqrt2\,g}\, \sum_{i=1}^4 q_i\, \vec
a_i\, X_i^{-1}\, b^{-2} +
\fft{d W}{d\vec \phi}\Big)\,,\nn\\
\fft{\dot b}{b} &=& -\fft{1}{2\sqrt2}\,
\Big(-\fft{\epsilon}{\sqrt{2}\,g}\,
\sum_{i=1}^4 q_i\, X_i^{-1}\, b^{-2} + W\Big)\,,\nn\\
\fft{\dot a}{a} &=& \fft{1}{2\sqrt2}\, \Big(-
\fft{\epsilon}{\sqrt2\,g}\, \sum_{i=1}^4 q_i\, X_i^{-1}\, b^{-2} -
W\Big)\,, \label{4}
\eea
provided that $\lambda_i=\epsilon\, g^{-1}\, q_i$, with $q_1 + q_2
+ q_3 + q_4=1$.   Here, the superpotential $W$ is given by
\be
 W=\sqrt2\, g\, \sum_{i=1}^4 X_i\,.
\ee
If all of the $q_i$ are equal, the scalars can be consistently set
to zero and the solution reduces to the previous four-dimensional
solution of section 2. In general, with an appropriate choice of
$q_i$, it is straightforward to find solutions that run from
dS$_2\times S^2$ in the infinite past to a dS$_4$-type geometry in
the infinite future.

\section{Conclusions}

     We have considered massive type IIA$^*$ theory in which the
kinetic terms for the R-R fields have the ``wrong'' sign. We find that
the theory admits a smooth solution as a warped product of dS$_6$ and
$H^4$. This enables us to perform a hyperbolic reduction to $D=6$,
${\cal N}=(1,1)$ pure gauged supergravity which admits dS$_6$
spacetime as its vacuum solution.  The gauge group is $SU(2)\times
U(1)$, which is the same as the $D=6$ AdS gauged
supergravity. However, the kinetic terms for the gauge fields have the
``wrong'' sign. This apparently undesirable feature enables us to
construct a cosmological solution that runs smoothly from the infinite
past, which is dS$_4\times S^2$, to the infinite future, with a
dS$_6$-type metric having an $R^3\times S^2$ boundary. One interesting
feature of the solution is that although it is time-dependent, it
arises from a first-order system {\it via} a superpotential
construction.  The solution provides an effective four-dimensional
cosmological model that incorporates de Sitter spacetime and also two
compact extra dimensions forming an $S^2$.  Of course since the solution
comes directly from string theory, the cosmological constant is
clearly too large in comparison to the observational data. It is
nevertheless interesting to observe that the * theories can provide
``supersymmetric'' time-dependent cosmological solutions.

      We applied the same analysis to $D=5$ and $D=4$, where the
theories are reductions of type IIB$^*$ and M$^*$-theory,
respectively.  We also provided the cosmological solutions for general
de Sitter Einstein-Maxwell theories.  It is of interest to note that
although AdS$_7$ gauged supergravity exists, the corresponding dS$_7$
theory does not.

       We also demonstrated that there is a larger class of
non-supersymmetric cosmological solutions arising from the
second-order equations of motion.  For an appropriate choice of
parameters, the solution can describe an expanding universe whose
expansion rate is significantly larger in the past than in the future,
providing a realistic inflationary model, with no singularity.  In
fact, with some appropriate matter coupling, we show that solutions
with such a property can also arise from first-order system.

      All of these cosmological solutions can be lifted to ten or
eleven dimensions.  In particular, we obtained two ways of smoothly
embedding dS$_4$ in massive type IIA$*$, with the internal space being
either $H^4\times S^2$ or an $H^4$ bundle over $S^2$.

      Although the * theories are necessary from the point of view of
time-like T-duality, they suffer from an instability due to the
ghost-like nature of the supergravity fields. It is of interest,
therefore, to study further whether the stability is protected by the
time-like T-duality or the ``supersymmetry,'' or whether the time
scale of the instability is large enough to nevertheless validate the
cosmological solutions.

\end{document}